\documentstyle[12pt]{article}

\newcommand{\be}{\begin{equation}}
\newcommand{\ee}{\end{equation}}
\newcommand{\bea}{\begin{eqnarray}}
\newcommand{\eea}{\end{eqnarray}}

\begin{document}

\begin{center}
\begin{large}
{\bf  A Note on Holography \\}
{\bf  on  a\\}
{\bf  Curved Brane  \\}
\end{large}  
\end{center}
\vspace*{0.50cm}
\begin{center}
{\sl by\\}
\vspace*{1.00cm}
{\bf A.J.M. Medved\\}
\vspace*{1.00cm}
{\sl
Department of Physics and Theoretical Physics Institute\\
University of Alberta\\
Edmonton, Canada T6G-2J1\\
{[e-mail: amedved@phys.ualberta.ca]}}\\
\end{center}
\bigskip\noindent
\begin{center}
\begin{large}
{\bf
ABSTRACT
}
\end{large}
\end{center}
\vspace*{0.50cm}
\par
\noindent

Some recent literature has examined the holographic-induced  cosmology of
 a brane universe in the background of an anti-de 
Sitter-black hole geometry. In this regard, curved-brane scenarios have
  begun to receive considerable attention.  
Our current interest is in
 a formal discrepancy that exists between  two such 
works by  Padilla (hep-th/0111247) and Youm (hep-th/0111276).
In particular, these authors have incorporated different values for
a conformal factor that  is used to relate the thermodynamics
of the relevant (AdS bulk and CFT brane) spacetimes.  
After a more general review,
we clarify  this issue and discuss  the implications on the
prior results.

\newpage

\par
Much of current gravitational physics  has been influenced by the
intriguing concept
of  ``holography'' \cite{tho,sus}. The underlying principle,
which has followed closely from the original 
``Bekenstein bound'' \cite{newbek}, is based on the notion
that the maximal  entropy 
that can be stored within a given volume will be
determined by the largest black hole 
fitting inside of that volume. Since  the entropy
of a black hole  is essentially given by its (outer-horizon) 
surface area \cite{bek,haw},
it follows directly that all relevant degrees of
freedom of {\bf any} system must (in some sense) live
on a  boundary enclosing that system.
\par
Since 't Hooft \cite{tho} and Susskind \cite{sus} originally proposed  
the holographic principle  as being fundamental to
any gravitational  theory (at least at a semi-classical level), it
has received its fair share of criticism. (See Ref.\cite{smolin}
for a review and references.) However, one undeniable success
of holography  
has been the celebrated duality that apparently
 exists between  anti-de Sitter (AdS) spacetimes 
and  conformal field theories (CFTs); that is,
the AdS/CFT correspondence \cite{mal,gub,wit}.
It has, in fact,  been  argued (and demonstrated in many instances)  that 
 the horizon thermodynamics  of a given  $n$+2-dimensional AdS
 black hole can be  identified with  a certain
 $n$+1-dimensional 
CFT.  
Furthermore,  the  CFT of interest is assumed
 to be in a strongly coupled, high-temperature regime and to live on a
timelike surface that can be identified as an asymptotic boundary
of the AdS spacetime.
\par
In a  paper of considerable interest \cite{ver}, Verlinde
applied the AdS/CFT correspondence (and other holographic considerations) 
to a  specific cosmological setting: an $n$+1-dimensional,
radiation-dominated, closed Friedmann-Robertson-Walker (FRW) universe.
In a subsequent study \cite{sav}, Savonije and Verlinde extended the
original treatment  so that the   
FRW universe is actually a CFT
living on the brane in an  Randall-Sundrum scenario 
\cite{ran}.\footnote{The
 Randall-Sundrum brane world \cite{ran} describes an $n$+1-dimensional 
submanifold (or $n$-brane)  that serves as a boundary for
an $n$+2-dimensional AdS bulk spacetime.  The basic premise (when $n=3$)
is that the ``ordinary'' matter of our universe is trapped on the
brane, whereas the graviton and (possibly) other hypothetical
fields are allowed to propagate in the  ``extra'' bulk dimension.
For further discussion and references, see Ref.\cite{rubakov}.}
In particular, the bulk spacetime
was regarded as an $n$+2-dimensional AdS-Schwarzschild black hole, with this
being
 bounded by an $n$+1-dimensional brane of constant tension.
Note that the brane tension is a free parameter
that was ultimately fixed to obtain a flat-brane solution.
\par
This pair of papers covered a wide scope, but let us summarize
some of the highlights (as relevant to the brane-world scenario). 
\par
(i) If one solves the field equations of the boundary action
(which effectively describes the brane dynamics),
 it leads to 
the standard Friedmann equations for a radiation-dominated universe
(assuming a suitable choice of brane tension).
In these equations, the ``Hubble constant'' ($H$) is identified with the 
time derivative
of  $a$; that is, the radial distance between the brane and the center of
the black hole. 
\par  
(ii) On the basis of the AdS/CFT correspondence,  the 
thermodynamics of the brane   can be directly
evaluated from the thermodynamic properties of the black hole
horizon. 
\par
(iii) The brane (i.e., CFT) entropy takes on a Cardy-like
form \cite{car}; with the Cardy ``central charge''
being directly related to the sub-extensive portion of
the energy (or entropy). This sub-extensive contribution can be identified
with the Casimir energy (or entropy) of the brane universe.  
\par
(iv) A special moment in the brane's  cosmological evolution is found
to be of particular significance: when the brane crosses the
black hole horizon.  At this point, the brane temperature
and entropy can be simply expressed in terms of the Hubble
parameter (and its derivative). Moreover, the first law
of brane thermodynamics coincides precisely with the
 Friedmann
equations. This is remarkable, inasmuch as 
the CFT  equations of state should, in principle, know nothing about gravity
or cosmology. 
\par
(v) When the brane crosses the horizon,  there are
further coincidences; including the Casimir entropy ($S_C$)
with the Bekenstein-Hawking entropy ($S_{BH}$) of a universal-size black
hole.\footnote{It has been shown  that, strictly speaking,
such coincidences are  
 not persistent when  quantum
effects have been accounted for  \cite{blah1}.  The interest of
the current letter, however, is restricted to the 
scenario where such quantum effects are essentially negligible.}
 If one follows Verlinde and conjectures that
$S_C\leq S_{BH}$ is actually a universal holographic bound
(which, in this case, is valid if the brane remains outside
of the horizon), then  various other coincidences can also
be interpreted as holographic bounds that are saturated
when the brane and horizon meet.  For instance,
given
a strongly self-gravitating Universe ($H^2a^2\geq 1$),
then the  CFT total entropy ($S$) is bounded by a quantity
known as the ``Hubble entropy'' ($S_H$). Meanwhile,
for a weakly self-gravitating universe ($H^2a^2\leq 1$),
$S$ is bounded by  a quantity referred to as the ``Bekenstein
entropy'' ($S_B$).  Note that the above terminology follows 
from the definitions of 
Ref.\cite{ver}.
\par
The Verlinde-Savonije program has since been extended
and generalized  to  various  AdS-black hole scenarios. (For a 
thorough list of the relevant citations,  consult  Ref.\cite{top}.)
More recently, there have been attempts at generalizing 
the formalism to de Sitter (dS) black holes \cite{noj4,blah2,new1,new2,new3,
top}.  Such attempts have had only qualified 
success, which can be attributed to  complications that are inherent
to any asymptotically  dS spacetime \cite{str2}. 
These complications have, as well, 
impeded the progress towards establishing  a dS/CFT
correspondence \cite{str}\footnote{For both earlier and later
work  with regard to this duality, we again refer the
reader to the bibliography of Ref.\cite{top}.} 
(at least to the level of its AdS analogue).
For instance,  dS spacetimes  lack a 
 globally timelike Killing vector and a spatial infinity (making it
difficult to define conserved charges), while
  the black hole horizon (and its  
thermodynamic properties)
 have an
ambiguous observer
dependence.
  It is also problematic
 that dS solutions are conspicuously absent in  string theories; thus  
inhibiting  rigorous  tests of the proposed duality.
\par
The investigations into dS holography have, in large part, been inspired
by observational data  that implies a positive cosmological constant
for the universe \cite{obs}.  In a brane-world scenario,
however, it is not clear what  properties should be attributed
to  the bulk spacetime.  On this basis and given that the AdS/CFT
duality is on firmer ground, an alternative viewpoint has  also
emerged \cite{wan3,blah3,ps,pad,youm}. 
 Namely, the bulk is treated as an AdS black hole,
but with an (effective) brane cosmological constant
that is non-vanishing and (for the dS case) 
positive.\footnote{In 
the Verlinde-Savonije treatment \cite{sav}, as well as  most subsequent
generalizations, the brane tension has been fixed
so that the (effective) brane cosmological constant is a vanishing
quantity.}
\par
In this letter, we are particularly interested in the last two 
 citations:
 Padilla \cite{pad} and Youm \cite{youm}. 
Both of these consider the just mentioned scenario of
a AdS-black hole bulk and a non-vanishing (positive) cosmological constant
on the brane.\footnote{Let us take note of  
 an earlier work
along these same lines by Nojiri and Odintsov \cite{blah3}. 
The focus of this study was more on the
implications of quantum effects, as opposed to the direct repercussions
of adding curvature to the brane. However, the  classical
results can still be obtained in the appropriate limits. We further
note that, in regard to the discrepancy discussed in the current letter,
Ref.\cite{blah3} shares the viewpoint of Youm \cite{youm}.}  
 There is, however, a slight discrepancy between
the two presentations, which should be clarified. Let us first discuss
the general set-up.
\par
The scenario of interest is an $n$+1-dimensional (single-sided) 
brane of constant
tension in an $n$+2-dimensional (topological) AdS-black hole 
bulk.\footnote{Note that we are specifically addressing the scenario
of Ref.\cite{youm}. On the other hand,  Ref.\cite{pad} considered a 
double-sided brane
and only ``pure'' AdS-Schwarzschild black holes. However, this
analysis can  be trivially  extended to the case we are considering.}
In a suitably static gauge, the bulk solution can be written as follows:
\be
ds^2_{n+2}=-h(a)dt^2+{1\over h(a)}da^2+a^2d\Omega^2_{n},
\label{1}
\ee
\be
h(a)=k+{a^2\over L^2}-{\omega_{n+1}M\over a^{n-1}},
\label{2}
\ee
\be
\omega_{n+1}={16\pi G_{n+2}\over n V_n}.
\label{3}
\ee
Here, $L$ is the curvature radius of the AdS background, $d\Omega_n^2$
denotes the line element of an $n$-dimensional
constant-curvature hypersurface
 with volume $V_{n}$, $G_{n+2}$ is
the $n$+2-dimensional Newton constant, and $M$ 
 and $k$  are constants of integration. $M$ is identifiable
as the ADM (i.e., conserved)
 mass of the black hole  and is always positive.
Meanwhile, without loss of generality, $k$ can be set to equal  +1, 0 or -1. 
These choices describe a  horizon
geometry that is respectively elliptic, flat or hyperbolic.
\par
The black hole horizon, $a_H$, is described by
the outermost root of the equation $h(a)=0$. Furthermore,
this horizon  has an associated temperature and entropy
that are respectively given as follows \cite{gh}:
\be
T_{AdS}={(n+1)a_{H}^2+(n-1)L^2 k\over 4\pi L^2 a_H} ,
\label{6}
\ee
\be
S_{AdS}={a_H^nV_n\over 4 G_{n+2}}.
\label{7}
\ee
 By virtue of the AdS/CFT correspondence \cite{mal,gub,wit}, it follows  
that the above thermodynamics
can be identified, up to a conformal factor, with the thermodynamics of
a CFT
that lives  on the brane. That is: 
\be
E\equiv E_{CFT}= {\cal C} M,
\label{18}
\ee
\be
T\equiv T_{CFT}={\cal C}T_{dS},
\label{19}
\ee
\be
S\equiv S_{CFT}= S_{dS}, 
\label{20}
\ee
where ${\cal C}$ is a yet unspecified conformal factor.\footnote{These
simple relations are known to be  violated  when, for instance,
 higher-derivative gravity terms are considered. See Ref.\cite{blah4}
for further discussion and references.}
 Note
that the entropy is {\bf not} subject to this rescaling \cite{wit}.
\par 
The dynamics of the brane can be determined by way of the following boundary
action:
\be
{\cal I}_b={1\over 8\pi G_{n+2}}\int _{\partial {\cal M}}
\sqrt{\left| g^{ind}\right|}{\cal K}
+{n\sigma\over 8\pi G_{n+2}}\int_{\partial {\cal M}}
\sqrt{\left| g^{ind}\right|},
\label{9}
\ee
where $g^{ind}_{ij}$ is the induced metric on the boundary  
($\partial{\cal  M}$),
${\cal K}\equiv {\cal K}^i_i$ is the trace of the extrinsic curvature
and $\sigma$ is a parameter measuring the brane tension.
Varying this action with respect to the induced metric, one obtains 
the following field equation:
\be
{\cal K}_{ij}={\sigma}g_{ij}^{ind}.
\label{10}
\ee
\par
In analogy with Ref.\cite{sav}, 
 the brane dynamics can be clarified with a 
new time parameter, $\tau$; whereby  $a=a(\tau)$,
$t=t(\tau)$ and:
\be
{1\over h(a)}\left({da\over d\tau}\right)^2-h(a)\left({dt\over
d\tau}\right)^2=-1.
\label{11}
\ee
\par
With  the above condition,
  the induced brane metric  adopts  an FRW form:
\be
ds^2_{n+1}=-d\tau^2+a^2(\tau)d\Omega^2_{n},
\label{12}
\ee
and the field equation (\ref{10}) translates to:
\be
{dt\over d\tau}={\sigma a\over  h(a)}.
\label{13}
\ee
\par
Defining the Hubble parameter in the usual way, $H\equiv {\dot a}/
a$,\footnote{A dot denotes differentiation with respect to $\tau$.}
we find that Eqs.(\ref{12},\ref{13}) yield the Friedmann equation
for radiative matter: 
\be
H^2={\omega_{n+1}M\over a^{n+1}}- {k\over a^2} +{2\over n (n-1)}\Lambda,
\label{14}
\ee
where $\Lambda={n (n-1)\over 2}\left(\sigma^2-{1\over L^2}\right)$
is the effective cosmological constant of the brane universe
(regarded as positive for the dS case). Unlike prior studies
(for instance, Ref.\cite{sav}), the brane tension is {\bf not}
fixed here  so that $\Lambda$ vanishes. 
\par
Also of interest, one obtains  the corresponding second Friedmann equation 
  by simply differentiating the above to give:
\be
{\dot H}= -{n+1\over 2}{\omega_{n+1}M\over a^{n+1}}+ {k\over a^2}.
\label{16}
\ee
\par
To proceed in the manner of Verlinde and Savonije, it is
necessary to identify the conformal factor, ${\cal C}$.
For a flat-brane scenario (i.e., $\sigma^2=L^{-2}$),  one simply
 rescales thermodynamic quantities by a factor
that corresponds to ${\dot t}$ in the $a\rightarrow\infty$ limit.
This leads to ${\cal C}= L/ a$. It is not quite as clear
what the analogous procedure should be when the brane has curvature.
In Ref.\cite{youm}, the author continues to use $L/ a$.
In Ref.\cite{pad}, however, it was argued that this
is not a consistent procedure. We discuss these arguments next.
\par
It is typically assumed that 
 the energy of the bulk directly 
corresponds  to the mass of the black hole.  
Padilla \cite{pad}, however, employed
a rigorous methodology to evaluate the bulk energy. 
This derivation was based on: (i) calculating the difference
in Euclidean actions (including  bulk and  boundary terms) 
between the AdS-black hole  and a  (pure) AdS reference geometry, 
(ii)  considering the high-energy limit  as is appropriate
for a dual CFT,\footnote{Besides high temperature, 
 various other justifiable assumptions were incorporated along the way.
See Ref.\cite{pad} for a complete discussion.} 
and (iii) using standard thermodynamics to extract the
energy from the Euclidean action \cite{gh,hh}.
The following result was obtained:
\be
E_{AdS}=\left({1\over L\sigma}\right)^2 M.
\label{blah}
\ee
Then, after asymptotically rescaling by ${\dot t}$ (see above), one finds:
\be
E_{CFT}= {L^2 \sigma\over a}E_{AdS}= {1\over \sigma a} M.
\label{blub}
\ee 
That is:
\be
 {\cal C}= {1\over\sigma a },
\label{hoop}
\ee
which does {\bf not} equal $L/a$ if the brane has a non-vanishing curvature.
\par
It is also significant to Padilla's approach that the
effective Newton constant on the brane ($G$) be
given as  follows \cite{pad2}:
\be
G= (n-1)\sigma G_{n+2}.
\label{huh}
\ee
This is contrary to the
 flat-brane expression of $G=(n-1) G_{n+2}/L$
\cite{ran}, which was also employed in Ref.\cite{youm}.  
\par
In Ref.\cite{youm}, unlike in Ref.\cite{pad},
Youm did a thorough analysis on  the cosmological
implications and holographic bounds of this  curved-brane 
scenario.\footnote{The author of 
Ref.\cite{pad} did, however, verify that the Friedmann equations
take on their anticipated forms for a radiation-dominated universe.}
The outcomes of Ref.\cite{youm} are, generally speaking, 
remarkably similar to those
 found for a flat brane  (see earlier discussion). One notable
exception is a failure in the CFT thermodynamics to reproduce
the first Friedmann equation when the brane crosses the horizon.
The author, however, explained this phenomenon:  the thermodynamic
properties of the CFT, which are determined from
the bulk black hole via the holographic principle, 
cannot know anything about the brane tension
(which  features prominently in the first Friedmann equation). Conversely,
the second Friedmann equation, which has nothing to say about
the brane tension, is indeed reproduced at the brane-horizon
coincidence point. 
\par  
One might wonder how the results of Ref.\cite{youm} 
would hold up  if Eqs.(\ref{hoop},\ref{huh}) are  incorporated
into the formalism.
As it happens, these two modifications conspire  against
each other so that  most of the  outcomes persevere unfettered.
This  relative invariance is (perhaps) indicative
of the ambiguity  in choosing  a  CFT dual to an AdS spacetime.
There are, however, some notable revisions that do indeed occur.
 We present these in the following itemized list. (For
calculational details, we defer to the  meticulous work 
of Ref.\cite{youm}.) 
\par
(i) Firstly, there is a correction  to the forms of the CFT entropy
and  temperature when the brane crosses the horizon:\footnote{The
original results will always be on the left and the modified
expressions, on the right.}
\be
S={1\over \sigma L}S_H \quad \rightarrow \quad S=S_{H},
\label{yipf}
\ee
\be
T=-\sigma L {{\dot H}\over 2\pi H} \quad \rightarrow \quad
T= -{{\dot H}\over 2\pi H},
\ee
where $S_{H}$ is the Hubble entropy as defined by  Verlinde \cite{ver}. 
Notably, the revised forms 
now agree with the flat-brane results. 
\par
(ii) Secondly, the  generalized Cardy formula should be  modified as follows:
\be
s^2=\left({4\pi \over n}\right)^2
\gamma \left[\rho-{k \gamma\over
a^2}\right] \quad \rightarrow \quad
s^2=\left({4\pi \sigma L\over n}\right)^2
\gamma \left[\rho-{k \gamma\over
a^2}\right]
\label{46}
\ee
where $s$ and $\rho$ are the  CFT entropy and energy density,  
and $\gamma$ (the Cardy central charge \cite{car}) 
is a  quantity  directly associated
with the Casimir energy on the brane. It would appear that
the CFT thermodynamics does know about the brane tension after all;  however,
not significantly, as  this factor of $\sigma$ is canceled
out by one in the temperature and does not appear in the 
first law of CFT thermodynamics. 
\par
(iii) Thirdly,  for a strongly self-gravitating universe (defined
by $S_B\geq S_{BH}$ \cite{ver}),\footnote{Here, $S_B$ are 
 $S_{BH}$ are Verlinde's definitions \cite{ver} of
the Bekenstein and  Bekenstein-Hawking entropy.
 When re-expressed in terms of
the Hubble constant, $S_B\geq S_{BH}$ does not simply translate
to $H^2a^2\geq 1$ as it does for a flat-brane scenario.
Rather, one now obtains: $H^2a^2\geq 2-k+{2a^2\over n(n-1)}\Lambda$.
 See Ref.\cite{youm} for further 
 details.} we obtain a modified form of the following holographic bound:
\be
S^2\leq {1\over \sigma^2 L^2}S_H^2 \quad\rightarrow\quad 
S^2\leq S_{H}^2.
\ee
Significantly, this revision agrees with the analogous flat-brane bound. 
\par
(iv) Finally, for a weakly self-gravitating universe ($S_B\leq S_{BH}$),
the  corresponding holographic bound should be corrected as follows:
\be
S\leq\sqrt{2-k}S_B \quad\rightarrow\quad
S\leq \sigma L\sqrt{2-k}S_B.
\label{yipl}
\ee
In this case,  the  revised and original forms both  fail to agree
with the analogous ($k=1$) flat-brane bound  of $S\leq S_B$. Hence,
the strongly self-gravitating entropic bound is (at least
in some superficial sense) more
universal than its weakly self-gravitating counterpart. This behavior
could not have been deduced from the  prior work of Ref.\cite{youm}. 
\par
In summary, we have considered  some recent papers \cite{pad, youm}  
that have  applied
 the holographic principle to an interesting scenario: an AdS-black hole bulk
with a dS solution on the brane. 
We began by discussing  the philosophy that underlies
such studies and reviewing the seminal (flat-brane) work \cite{sav}.
After  introducing the model of interest, we  
  went on to clarify a formal
 discrepancy that exists between  Ref.\cite{pad} and Ref.\cite{youm}.
This discrepancy involved a conformal factor that
is used to relate the dual (AdS bulk and CFT brane) spacetimes.
Finally, we have considered the implications of this distinction on the 
prior results
and documented the few  modifications of significance: 
Eqs.(\ref{yipf}-\ref{yipl}). That such corrections are essentially
trivial  probably reflects the (limited) freedom  one has in assigning
a dual CFT to a bulk AdS spacetime.


\begin{thebibliography}{99}

\bibitem{tho} G. 't Hooft, ``Dimensional Reduction in Quantum Gravity'',
gr-qc/9310026 (1993).
\bibitem{sus} L. Susskind, J. Math. Phys. {\bf 36}, 6377 (1995) 
[hep-th/9409089].
\bibitem{newbek} J.D. Bekenstein, Phys. Rev. {\bf D23}, 287 (1981).
\bibitem{bek} J.D. Bekenstein, Lett. Nuovo. Cim. {\bf 4}, 737 (1972);
Phys. Rev. {\bf D7}, 2333 (1973); Phys. Rev. {\bf D9}, 3292 (1974).
\bibitem{haw} S.W. Hawking, Comm. Math. Phys. {\bf 25}, 152 (1972);
 J.M. Bardeen, B. Carter and S.W. Hawking, Comm. Math. Phys.
{\bf 31}, 161 (1973).
\bibitem{smolin} L. Smolin, Nucl. Phys. {\bf B601}, 209 (2001)
[hep-th/0003056]. 
\bibitem{mal} J.M. Maldacena, Adv. Theor. Math. Phys. {\bf 2},
231 (1998) [hep-th/9711200].
\bibitem{gub} S.S. Gubser, I.R.  Klebanov and A.M. Polyakov,
Phys. Lett. {\bf B428}, 105 (1998)
 [hep-th/9802109].
\bibitem{wit} E. Witten, Adv. Theor. Math. Phys. {\bf 2}, 253 (1998)
[hep-th/9802150].
\bibitem{ver} E. Verlinde, ``On the Holographic Principle in a
Radiation Dominated Universe'', hep-th/0008140 (2000).
\bibitem{sav} I. Savonije and E. Verlinde, Phys. Lett. {\bf B507},
305 (2001) [hep-th/0102042].
\bibitem{ran} L. Randall and R. Sundrum, Phys. Rev. Lett. {\bf 83},
3370 (1999) [hep-ph/9905221]; {\it ibid}, 4690 (1999)
[hep-th/9906064].
\bibitem{rubakov} V.A. Rubakov, ``Large and Infinite Extra
Dimensions'', hep-ph/0104152 (2001).
\bibitem{car} J.L. Cardy, Nucl. Phys. {\bf B270}, 317 (1986).
\bibitem{blah1} S. Nojiri and S.D. Odintsov, Int. J. Mod. Phys.
{\bf A16}, 3273 (2001) [hep-th/0011115].
\bibitem{top} A.J.M. Medved,  
``CFT on the Brane with a Reissner-Nordstrom-de Sitter
Twist'', hep-th/0111182 (2001);
``dS/CFT Duality on the Brane with
a Toplogical Twist'', hep-th/0111238 (2001).
\bibitem{noj4} S. Nojiri, O.  Obregon, S.D.  Odintsov, H. Quevedo and M. P.
 Ryan, Mod. Phys. Lett. {\bf A16}, 1181 (2001) [hep-th/0105052].
\bibitem{blah2} S. Nojiri and S.D. Odintsov, ``Quantum Cosmology, Inflationary
Brane-World Creation and dS/CFT Correspondence'', hep-th/0107134
(2001).
\bibitem{new1} U.H.  Danielsson, 
``A Black Hole Hologram in de Sitter Space'',
 hep-th/0110265 (2001).
\bibitem{new2}  S. Ogushi, ``Holographic Entropy on the
Brane in de Sitter Schwarzschild Space'', hep-th/0111008 (2001).
\bibitem{new3}  R.-G.  Cai, ``Cardy-Verlinde Formula
and Asymptotically de Sitter Spaces'',
 hep-th/0111093  (2001).
\bibitem{str2} For a general discussion on de Sitter space:
 M.  Spradlin, A. Strominger and  A. Volovich,
``Les Houches Lectures on De Sitter Space'', 
hep-th/0110007 (2001).
\bibitem{str} A. Strominger, JHEP {\bf 0110}, 034 (2001) [hep-th/0106113].
\bibitem{obs} For instance, see:  N. Bahcall, J.P. Ostriker, S. Perlmutter
and P.J. Steinhardt, Science {\bf 284}, 1481 (1999)
[astro-ph/9812133].
\bibitem{wan3}  B. Wang, E. Abdalla and R.-K.  Su,
Phys. Lett. {\bf B503}, 394 (2001) [hep-th/0101073]; ``Friedmann
Equation and Cardy Formula Correspondence in Brane Universes'',
hep-th/0106086 (2001).
\bibitem{blah3} S. Nojiri and S.D. Odintsov, Class. Quant. Grav. {\bf 18},
5227 (2001) [hep-th/0103078].
\bibitem{ps} A.C. Petkou and G. Siopsis, ``dS/CFT Correspondence
on a Brane'', hep-th/0111085 (2001).
\bibitem{pad} A. Padilla, ``CFTs on Non-Critical Braneworlds'',
hep-th/0111247 (2001).
\bibitem{youm} D. Youm, ``The Cardy-Verlinde Formula and
Asymptotically de Sitter Brane Universe'', hep-th/0111276
(2001).
\bibitem{gh} G.W. Gibbons and S.W. Hawking, Phys. Rev. {\bf D15},
2752 (1977).
\bibitem{blah4} S. Ogushi, ``AdS Black Hole in R$^2$ Gravity'',
hep-th/0108209 (2001).
\bibitem{hh} S.W. Hawking and G.T. Horowitz, Class. Quant. Grav.
{\bf 13}, 1487 (1996) [gr-qc/9501014].
\bibitem{pad2} R. Gregory and A. Padilla, ``Nested Braneworlds and
Strong Brane Gravity'', hep-th/0104262 (2001).





\end{thebibliography}
\end{document}